# BAUM: A DNA ASSEMBLER BY ADAPTIVE UNIQUE MAPPING AND LOCAL OVERLAP-LAYOUT-CONSENSUS

Anqi Wang[1], Zheng Li[1], Zhanyu Wang[1] and Lei M. Li[1,*]

[1]: National Center of Mathematics and Interdisciplinary Sciences, Academy of Mathematics and Systems Science, Chinese Academy of Sciences, Beijing, 100190, China

[*]: Correspondence should be addressed to Lei M. Li (lilei@amss.ac.cn). Telephone: +8610-82541585.

**Running title: Assembler by Adaptive Unique Mapping and Local OLC**

# ABSTRACT

Genome assembly from the high-throughput sequencing (HTS) reads is a fundamental yet challenging computational problem. An intrinsic challenge is the uncertainty caused by the widespread repetitive elements. Here we get around the uncertainty using the notion of uniquely mapped (UM) reads, which motivated the design of a new assembler BAUM. It mainly consists of two types of iterations. The first type of iterations constructs initial contigs from a reference, say a genome of a species that could be quite distant, by adaptive read mapping, filtration by the reference's unique regions, and reference updating. A statistical test is proposed to split the layouts at possible structural variation sites. The second type of iterations includes mapping, scaffolding/contig-extension, and contig merging. We extend each contig by locally assembling the reads whose mates are uniquely mapped to an end of the contig. Instead of the de Bruijn graph method, we take the overlap-layout-consensus (OLC) paradigm. The OLC is implemented by parallel computation, and has linear complexity with respect to the number of contigs. The adjacent extended contigs are merged if their alignment is confirmed by the adjusted gap distance. Throughout the assembling, the mapping criterion is selected by probabilistic calculations. These innovations can be used complementary to the existing *de novo* assemblers. Applying this novel method to the assembly of wild rice *Oryza longistaminata* genome, we achieved much improved contig N50, 18.8k, compared with other assemblers. The assembly was further validated by contigs constructed from an independent library of long 454 reads.

# INTRODUCTION

Genome assembly from high-throughput sequencing (HTS) reads is a fundamental yet challenging computational problem in the genome research. Two major frameworks of assembly methods are the overlap-layout-consensus (OLC) paradigm (Staden 1980) and the de Bruijn graph (DBG) representation (Waterman 1995; Pevzner et al. 2001) of *k*-mers. Although both frameworks were proposed during the era of genome assembly based on Sanger sequencing reads, OLC played a more important role before the coming of HTS technology. The complexity of OLC is quadratic with respect to the number of reads and it is difficult to implement OLC assemblers to the whole set of HTS reads from a sequencing project. Consequently, the DBG model became a practical and popular choice for DNA assembly. An intrinsic challenge faced by any assembler is the uncertainty of read origins caused by widespread repetitive elements across a genome. The degree of uncertainty is particularly high for the HTS short reads.

Motivated by these considerations, in this article we propose an assembler by adaptive unique mapping (BAUM) and local overlap-layout-consensus. We get around the uncertainty using the notion of uniquely mapped (UM) reads defined by a mapping criterion. Unique mapping is used to generate layouts, to construct scaffolds and to extend contigs, and thus the mapping criterion needs to be set adaptively. Assisted by a reference, say a draft genome of a species that could be quite distant, the initial contigs are obtained by the layouts of UM reads, a filtration, and layout split at the sites of possible structural variations. By integrating UM and paired-end/mate-pair information, we stratify reads, mainly outside the current layouts, into groups, which can be assembled locally by OLC to extend the contig in a parallel way. The complexity of the local OLC is quadratic only with respect to the read depth and is linear with respect to the number of contigs. The adjacent extended contigs are merged if their alignment is confirmed by statistically adjusted gap distance. The result in turn can serve as the basis for the next round of mapping-scaffolding-extension. Throughout the procedure, we provide statistical measures for the mapping criterion selection and for the layout split.

# RESULTS

The sequencing data considered in this article are from HTS technology with paired-end/mate-pair libraries. The scheme of the assembler BAUM is shown in **Figure 1**. It mainly consists of two parts shown respectively by a loop in red on the left and a loop in green on the right.

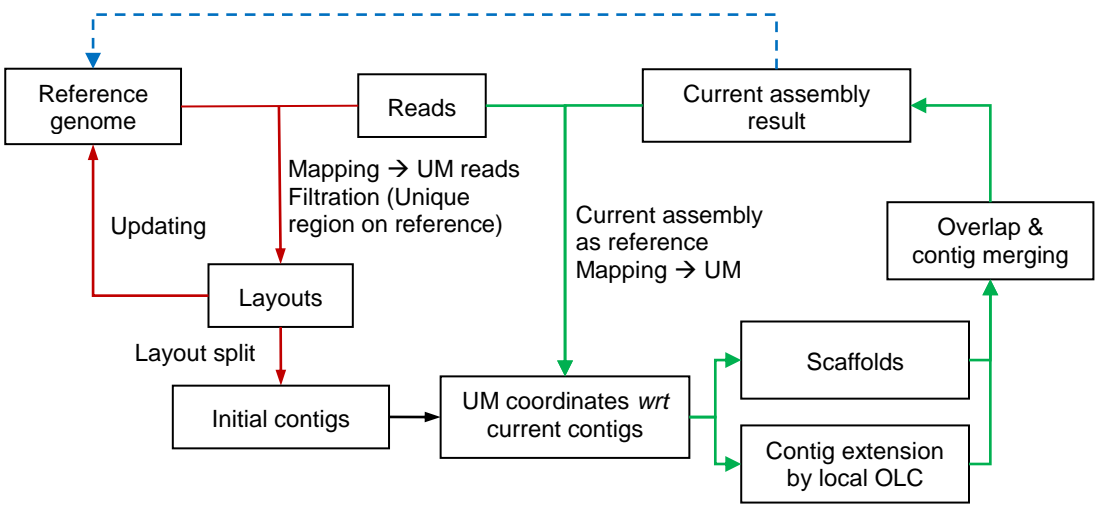

**Figure 1: The scheme of the assembler BAUM.** BAUM mainly consists of two loops shown respectively in red and in green. The first loop constructs the initial contigs by the UM reads with respect to a reference. At the beginning of the loop, reads are mapped to the reference genome and layouts of the UM reads are generated. The layouts falling out of the reference's unique-regions are filtered out. The reference's unique-regions are also defined by UM reads that are obtained by mapping all pseudo-reads generated from the reference to itself. To increase the mapping rate, we update the reference's unique-regions by the consensus of the layouts, and re-map the reads according to a new UM criterion adaptively. After a step of layout split, the obtained initial contigs are set to be the current contigs, and enter the second loop. In the second loop, scaffolds are built based on paired-end/mate-pair information, and those reads whose mates are uniquely mapped to an end of a contig, are locally assembled using the overlap-layout-consensus (OLC) paradigm to extend the contig. The adjacent extended contigs are merged if they overlap in consistency with the adjusted gap distance. The resulting assembly in turn can serve as the basis for the next round of mapping-scaffolding/extension-merging.

The first loop on the left shows the construction of initial contigs by the UM reads with respect to a reference, a draft genome of a species that could be quite distant. At the beginning of the loop, reads are mapped to the reference genome and layouts of the UM reads are generated. The layouts falling out of the reference's unique-regions are filtered out. The reference's unique-regions are defined by UM reads that are obtained by mapping all pseudo-reads generated from the reference to itself. If the divergence between the target and the reference genome is relatively large, then we update the reference's unique-regions by the consensus of the layouts, and re-map the reads according to a new UM criterion adaptively.

However, if a relatively large structural variation such as insertion/deletion occurs at a site on the target genome with respect to the reference, we need to make sure the layout of UM reads is split at this site. A statistical test is proposed for the layout split. Its primary goal is to reduce the error that a necessary split is missed. After layout split, the initial contigs are set to be the current contigs, and enter the second loop.

In the second loop on the right in **Figure 1**, scaffolds are built based on the mate-pair information. Meanwhile, those reads whose mates are uniquely mapped to an end of a contig, are locally assembled using the overlap-layout-consensus (OLC) paradigm to extend the contig. The consensus sequences of every two adjacent extended contigs on the scaffolds are merged if they are found overlap. The merging criterion involves the consistency examination of the gap distance. After contig merging, we obtain the current assembly. The resulting current assembly can be taken as a new reference genome for the next iteration of assembly in which the mapping, scaffolding/contig extension, and merging are carried out again. The contig N50 is expected to be improved after each iteration. We note that the mapping criterion is selected adaptively at each iteration. In the METHODS section, we explain the key steps in details.

**Assessment of initial contigs by simulations.** The validity of the initial contigs, which represent the genome regions without uncertainty, is crucial for the success of the proposed assembler. Other than the analytical results, we assessed their reliability by a simulation study. Reads of 90bp long and 150× were uniformly generated from the published genome of *Oryza nivara* (OMAP 2014). Assisted by the reference of *Oryza sativa japonica* (IRGSP-1.0), we built the initial contigs by BAUM. The parameters $k$ and $m$, which were respectively the lower bound of perfect match seed length and the upper bound of mismatches, in the mapping algorithm SEME (Chen et al. 2013) were taken to be 23 and 4 respectively. The threshold for the depth in filtration was 250. In the layout split, both the parameters $W_1$ and $W_2$, defined as the maximum tail length of the leftward reads and rightward reads at a base on the reference genome(details are explained in the METHODS section), were set to be 40. The resulting initial contigs were then aligned to the chromosomes of *Oryza nivara* using BLAST+ (Camacho et al. 2009) and the proportion of perfect alignment is 0.997. Thus the assembly error at the structural level was well controlled by the unique mapping, filtration and layout split.

**Assembly of *Oryza longistaminata*.** To further test the performance of the proposed method, we used BAUM to assemble the genome of *Oryza longistaminata*, a wild rice from Africa. The sequencing data included seven Illumina HiSeq paired-end/mate-pair libraries (insert sizes are 300, 400, 1K, 2K, 5K, 10K and 20K) with a total sequencing depth 230× and the read lengths were between 90bp and 100bp. As reported by Zhang et al. (2015), the initial contig N50 was less than 1k when applying SOAPdenovo2 (Luo et al. 2012) to the HiSeq data. Then additional sequencing data of 454 long reads were obtained and were integrated with the contigs generated by SOAPdenovo2. The final contig N50 and scaffold N50 were 12.5k and 363k respectively, and the genome size was 342M. The estimated genome size based on the cytometry test is 329M.

We carried out our assembly workflow using only the HiSeq data. A kind of cultivated rice, *Oryza sativa japonica* (IRGSP-1.0), was chosen as our initial reference. The *longistaminata* and the *japonica* have potentially diverged from *Oryza glaberrima* about 1.9 and 0.6 million years ago respectively (Zhang et al. 2015).

**First stage: updating reference.** The divergence between *japonica* and *longistaminata* is fairly large, and we therefore ran four rounds of read-mapping and reference-updating using the paired-end libraries of insert size 300, 400 and 1K (parameters and results are shown in **Table 1(a)**). In each round, reads were mapped to the updated reference genome using SEME. The perfect seed length $k$ was set 23 in all rounds, while the upper bound of mismatches $m$ was set 20, 15, 13 and 10 respectively. As shown in the left loop of **Figure 1**, we updated the reference using the layouts of UM reads. Taking a single nucleotide variation for example, we updated the base at the site by the most frequent nucleotide if its frequency exceeded a certain threshold, and the thresholds in the four rounds were set as 90%, 85%, 80% and 75% respectively. It can be seen from **Table 1(a)** that the mapping criteria in the updating process got more and more stringent, while the mapping rate of each library remained at the same level along the updating process. This indicated that the reference was updated effectively. After reference-updating, reads were mapped to the last updated reference by setting $k$ as 23 and $m$ as 6. The proportion of the multiply mapped reads in the successfully mapped reads was 21%, representing the proportion of uncertainty due to repetitive elements. Through self-mapping pseudo-reads under the same criterion, a total length of 134Mb were defined as non-unique regions, which accounted for about 1/3 of the reference *japonica* genome. Layouts falling into these non-unique regions were then filtered out. Since the layouts were short due to the fairly large divergence between the two genomes, we did not impose stringent filtration on the depth. In the layout split, the threshold for the hypothesis tests was set 15 so that the false positive rate can be controlled under 0.0067 (see Proposition 2 in METHODS for details). The N50 of the initial contigs was 1,081bp.

**Table 1**: Results of *Oryza longistaminata*'s assembly obtained by BAUM

(The last two columns in (b) list the unique and multiple mapping rates of the 300 bp library.)

(a) The first stage: updating

| Stage I | Upper bound of mismatch number | Threshold of nucleotide frequency (%) | Mapping rate for paired-end libraries | | |
|---|---|---|---|---|---|
| | | | 300 bp (%) | 400 bp (%) | 1 kbp (%) |
| A | 20 | 90 | 83.8 | 83.1 | 72.9 |
| B | 15 | 85 | 84.2 | 83.2 | 73.4 |
| C | 13 | 80 | 83.9 | 82.8 | 73.1 |
| D | 10 | 75 | 82.7 | 81.3 | 71.9 |

(b) The second and third stages

|  | Contig N50 (bp) | Scaffold N50 (bp) | Total size (Mbp) | Unique mapping (%) | Multiple mapping (%) |
|---|---|---|---|---|---|
| **Stage II** | | | | | |
| Pre-assembly | 21,220 | 334,726 | 339 | 73.0 | 13.00 |
| **Stage III** | | | | | |
| Initial contigs | 2,592 | --- | 233 | 55.4 | 0.52 |
| Iteration 1 | 10,390 | 624,458 | 309 | 66.1 | 5.46 |
| Iteration 2 | 15,536 | 535,614 | 299 | 68.8 | 4.73 |
| Iteration 3 | 18,203 | 459,253 | 303 | 71.2 | 6.54 |
| Scaffolds longer than 1Kbp | 18,782 | 481,584 | 296 | 71.4 | 6.06 |

**Second stage: pre-assembly and updated initial contigs.** After obtaining the initial contigs, we carried out four rounds of pre-assembly as shown in the right loop of **Figure 1**. During the process, we constructed scaffolds by SSPACE (Boetzer et al. 2011) in the first and second iteration. SSPACE was run under the TAB mode to avoid repeated mapping. The upper bounds of mismatches for mapping in the last three iterations were set as 6, 5 and 4 respectively, which got more and more stringent. The three paired-end libraries, 300, 400, and 1K, were used for contig extension in the iterative assembly, while the 1K library was skipped in the first-round assembly because the contig N50 was small. Then we applied Smith-Waterman algorithm to detect any possible overlapping of every two adjacent contigs. The score for match was set as 1, and the scores for mismatch, gap open and gap extension were all set as -3. We merged the adjacent extended contigs for all the alignments with score greater than 20. This merging strategy resulted in a pre-assembly in which false negatives are well controlled. The false positive merging would be eliminated by another round of layout split shown in the left loop.

We took the pre-assembly as the reference (shown as the dashed blue arrow in **Figure 1**) and ran the mapping, filtration and layout split again. Since the reference genome in the second stage was much closer to the target genome, we adopted more stringent rules including: (1) setting the upper bound of mismatch number in the initial mapping as 3; (2) carrying out the filtration by setting the depth threshold as 330 (the mode of the depths was 180); (3) increasing the threshold in the hypothesis test for the layout split to 30 (the theoretical false positive is $9.6 \times 10^{-14}$; see Proposition 2 in METHODS for details). The N50 of the updated initial contigs was 2.6k.

**Third stage: final assembly.** Based on this new initial contigs, we ran three iterations of the right loop in the BAUM pipeline, namely, mapping, contig extension/scaffolding, and contig merging. The maximum mismatch number in the mapping was set as 2. Again the scaffolds were generated by SSPACE, in which the lower bound of link number -k was set as 15. The contigs were extended along both ends by PHRAP (Green et al. 1999) in a parallel configuration. The contig-extensions in all iterations including those in the pre-assembly stage were completed within three hours on a Dell XD720 server with 24 CPU cores and 384GB memory.

Unlike the line of "merging as much as possible" in the previous stage, we only merged the adjacent contigs if they were found overlap with additional confirmation. The confirmation came from an independent gap-size estimation resulted from SSPACE. Next we explain the details.

The first step is still the application of the local alignment (Waterman 1995) to the adjacent extended contigs. The scores were set the same as the previous stage, and those alignments with scores greater than 20 were collected as candidates for contig merging. As shown in **Figure 2**, the candidate alignments included three types, (a) no hanging end, (b) one hanging end and (c) two hanging ends. It would cause false positives to merge them all without additional confirmation. To reduce false positives, we resorted to an independent source of information which were the gap distances estimated by SSPACE using the insert sizes of libraries. Nevertheless, these estimates by SSPACE were not perfect. At this point, we had two gap distance estimates for each adjacent extended contig pair: one from SSPACE, which we referred to as *insert distance*, the other from local alignment, which we referred to as *alignment distance*. Our approach to the selection of contig merging was the following: we only merged those pairs of adjacent contigs whose insert distance and alignment distance were consistent. Technically, we examined the consistency by running a simple linear regression of the alignment distances with respect to the insert distances. Those pairs whose residuals were beyond a threshold were not merged. We first considered the "no hanging end" occasion, whose scatter plot was shown in **Figure 2(d)**. In order to remove the effect of outliers, we applied the least trimmed squares regression (LTS) (Rousseeuw and Leroy 1987; Li 2005), a robust statistical method that minimizes the sum of squared residuals among all subsets of a given size, to estimate the parameters of the linear model. Namely, for a fixed $k$, the objective function that LTS minimizes is $S_k(\beta) = \sum_{i=1}^{k} r_{(i)}(\beta)^2$, where $r_{(i)}(\beta)$ denotes the $i$-th ordered absolute values of the residuals (in the increasing order). To a great extent, the parameter $k$ controls the robustness of LTS against outliers. In this case of **Figure 2(a)**, the LTS selected a subset containing 91% points of the whole data and their correlation was 0.94, which was substantially greater than the correlation 0.58 for the entire data. The regression line based on this selected subset was: $alignment\ distance = 1.015\ insert\ distance - 89.13$. The $R^2$ of this LTS linear model was 0.89, much higher than 0.33 from the standard least squares method. The slope of regression line was close to the expected value 1, compared with the slope 0.36 from the standard least squares method. The intercept suggested that the insert distances obtained by SSPACE had some bias. To measure the variability of the difference between the two distances, we also used a robust estimator, median absolute deviation (MAD). Since $\sigma \approx 1.4826\ MAD$ for normally distributed data, we set a threshold 200 (about $3\sigma$) to determine the confirmation from alignment distance to insert distance. Consequently we applied an acceptance region to merge: $|insert\ distance - alignment\ distance - 89| < 200$. We applied the same regression model to the other two types of alignment shown in **Figure 2(b, c)**. In **Figure 2(d, e, f)**, the solid lines refer to the regression line and the dashed lines refer to the thresholds. The blue points correspond to the merged contig pairs and the red ones correspond to the unmerged contig pairs.

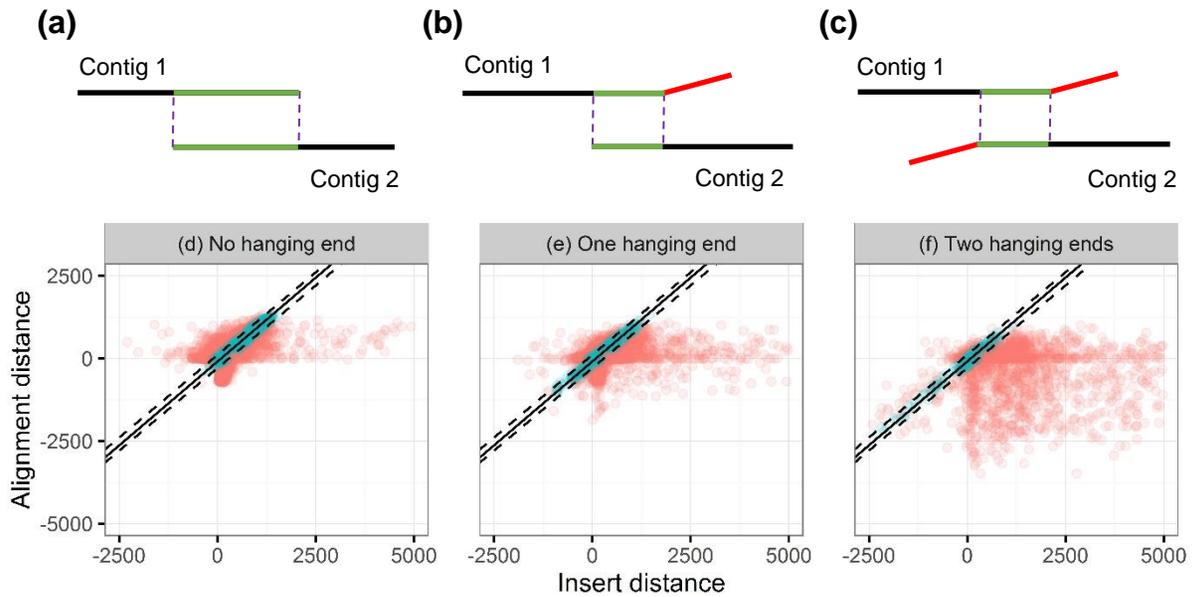

**Figure 2: Three alignment types and selective contig merging.** We applied the local alignment to adjacent contigs on the scaffolds. The candidate alignments included three types, (a) no hanging end, (b) one hanging end and (c) two hanging ends. To reduce false positives of contig merging, we check the consistency between two gap distance estimates: the *alignment distance* estimated by local alignment and the *insert distance* estimated from insert sizes by SSPACE. Technically, we examined the consistency by running a simple linear regression of the alignment distances with respect to the insert distances. Those pairs whose corresponding residuals were beyond a threshold were not merged. In order to remove the effect of outliers in the parameter estimation, we applied the approach of least trimmed squares (LTS). In this case

of **Figure 2(a)**, the LTS selected a subset containing 91% points of the data. The correlation increased substantially from 0.58 for the entire data set to 0.94 for the LTS subset. The resulting linear regression was: $alignment\ distance = 1.015\ insert\ distance - 89.13$. Now the slope was close to the expected value 1, and the intercept suggested that the insert distances obtained by SSPACE had some bias. We used the median absolute deviation (MAD) to measure the variability of the difference between the two distances. As $\sigma \approx 1.4826 MAD$ for normally distributed data, we set a threshold 200 (near $3\sigma$) to accept the consistency between alignment distance and insert distance. We applied the same regression model to the other two types of alignment shown in **Figure 2(b, c)**. The points corresponding to merged contigs in **Figure 2(d, e, f)** are marked in blue and others are marked in red.

Shown in **Table 1(b)** are the results of each step in the second and third stage. The final size of the assembled genome was 296M; the contig and scaffold N50 were 18.8K and 482K respectively. The last two columns listed the unique mapping rate and multiple mapping rate of the insert size 300bp library against the current assembly in each step. Considering the genome has a high heterozygosity rate (Zhang et al. 2015), we set the mismatch bound to be 7 in the mapping criterion. The multiple mapping rate of initial contigs in Stage III was 0.52%, which indicated that the non-unique regions or repetitive elements were by and large absent in the initial contigs. In other words, the majority of the initial contigs were unique-regions as defined by a certain mapping criterion, and this is consistent with the design of the assembler BAUM.

**Comparisons with other assemblers.** We compared BAUM with SOAPdenovo2, AbySS (Simpson et al. 2009) and Allpaths-LG (Gnerre et al. 2011) based on the *Oryza longistaminata* sequencing data. For each of the three assemblers, we tried it under different parameters and selected the best result. The running of SOAPdenovo2 included the process of gap closing. For the Allpaths-LG, two assembly results were obtained: one under the standard mode and the other under the cheat mode, in which the genome of *japonica* was used to assist the assembly. We only used the 2Kbp mate-pair library in scaffolding for Allpaths-LG, because the Allpaths-LG crashed down when more than one mate-pair libraries were incorporated. Since the Allpaths-LG did not output the scaffolds longer than 1kbp, we deleted the scaffolds shorter than 1K in the other assembly results for fair comparisons. **Table 2** lists the results of all assemblers regarding contig N50, scaffold number, scaffold N50, total size, non-N base size, unique mapping rate and multiple mapping rate of the 300bp paired-end library. The mapping rates were calculated by SEME and the mismatch bound in the mapping criterion was set as 7. It can be seen that BAUM had significant improvements over the other assemblers in terms of contig N50 and scaffold N50. The contig and scaffold N50 were even better than the hybrid assembly reported in (Zhang et al. 2015), which integrated the Illumina sequencing reads and 454 sequencing reads. After the deletion of the scaffolds shorter than 1Kbp, the total size of the BAUM assembly was still close to the estimated value given by the cytometry test, while those corresponding to SOAPdenovo2 and Allpaths-LG were obviously smaller; the non-N base numbers of the three assemblers were only around 200M. It is noticed that ABySS got larger total size and non-N base size than BAUM, but the mapping rates for ABySS were even lower. We also observed that the assistance of the *japonica* genome made little improvement on the assembly for Allpaths-LG. This is reasonable since the divergence between the two genomes was fairly large. The BAUM iterations, however, converged to the target genome from the *japonica* genome through multiple rounds of adaptive UM while detected structural variations by layout splits.

**Table 2**: Comparison of BAUM with other assemblers

| Assembler | Contig N50 (bp) | Scaffold Number | Scaffold N50 (bp) | Total size (Mbp) | Non-N size (Mbp) | Unique mapping (%) | Multiple mapping (%) |
|---|---|---|---|---|---|---|---|
| SOAPdenovo2 | 5,446 | 12,090 | 160,995 | 279 | 185 | 48.6 | 1.17 |
| ABySS | 3,313 | 48,617 | 14,848 | 320 | 299 | 66.7 | 4.98 |
| Allpaths-LG | 8,120 | 18,010 | 27,747 | 228 | 207 | 61.2 | 4.21 |
| Allpaths-LG (c)* | 8,056 | 18,112 | 27,197 | 228 | 206 | 61.0 | 4.19 |
| BAUM | 18,782 | 7,684 | 481,584 | 296 | 270 | 71.4 | 6.06 |

*: (c) indicates the cheat mode, in which the genome of *japonica* was used to assist the assembly.

Note: Only the 2Kbp mate-pair library was used in scaffolding for Allpaths-LG and Allpaths-LG (c). Scaffolds shorter than 1Kbp were filtered out in all the assembly results.

**Validation by 454 assembly.** We examined the correctness of the BAUM-Illumina results with the help of the independent library of 454 long reads, which were initially generated for the purpose of improving the contig N50 (Zhang et al. 2015). They were about 700bp long, and the coverage was 5.93. We assembled the 454 long reads using Newbler (454-Life-Sciences 2012). For each long 454 contig, we aligned it to the BAUM-Illumina scaffolds using BLAST+ and selected the longest HSP (High-scoring Segment Pair). Among the top 100 longest 454 contigs whose lengths were beyond 17K, the average length of the longest HSPs was 17,283 and the average identity was 99.57% (details are listed in Supplemental Table S1).

Overall, the contig N50 of the BAUM-Illumina assembly was substantially better than the hybrid assembly (Zhang et al. 2015) and the other assemblers listed above. Meanwhile, the false positives of the contig continuity were well controlled by the selection of mapping criterion and the splits of layouts.

## DISCUSSION

In this article, we propose a genome assembly method BAUM that is entirely different from the mainstream Eulerian path methods which are based on the representation of reads by the de Bruijn graph. BAUM does not involve the decomposition of the read into $k$-mers at all. The method does need a reference as a starting point. A close reference genome is good while a relatively distant genome also works as well as shown in RESULTS. In the rice example, *Oryza longistaminata* and *Oryza sativa japonica* have potentially diverged from *Oryza glaberrima* ~1.9 and ~0.6 million years ago (Zhang et al. 2015). In the case without any reference, we can start with a draft genome obtained by a DBG method. In this sense, the current method is complementary to the existing algorithms.

Both computationally and statistically, BAUM relies on mapping of reads to a reference genome. In fact, any efficient mapping tools for HTS reads can be used. We adopted SEME in the examples because SEME is accompanied with a quantitative evaluation that can guide the selection of UM criterion. Moreover, due to the algorithm design, SEME is still efficient when a relatively large number of mismatches are allowed in the alignment. This is crucial to BAUM since we need to update the reference genome for several rounds when the divergence between the reference and target genome is fairly large.

The concept of unique regions has been used in the DNA assembly in the era of Sanger sequencing, whose read length can reach as long as 1000bp, see Celera (Myers et al. 2000) and ARACHNE (Batzoglou et al. 2002). In the context of HTS, the read lengths are short and the balance between the couple of uncertainty and uniqueness is different. BAUM uses UM reads to deal with uncertainty. The unique regions on the reference genome are defined by the UM pseudo-reads generated from the reference itself. The criterion of UM is governed by the specificity of mapping, which under a random setting can be evaluated quantitatively as in Proposition 1. In different stages of the assembling, BAUM takes different mapping criteria according to the difference between the reference and the target genome, and consequently the uniqueness is defined adaptively. In comparison, the mapping criterion is usually fixed in other mapping based methods.

The layouts aligned to the unique regions directly generate the initial contigs, the basis of scaffolding/contig extension and merging. However, before moving forward, we need to split the layouts at any sites of structural variations (SV) between the reference and the target genome. Most existing methods detect SVs by checking the read depths, namely, from a "vertical" view. BAUM divides the reads covering a position into leftward and rightward reads, and considers their "horizontal" information as well. The false negatives are controlled by a setup of statistical testing, and are also validated by the simulation in RESULTS. After layout split, we need to construct scaffolds by the initial contigs and to close the gaps if possible. In the literature a closely related topic is referred to closure of gaps within scaffolds using paired-end reads. One such example is GapFiller (Boetzer and Pirovano 2012), which selects reads using the criteria: a) one read can be aligned to an existing contig; and b) the mate of the read partially falls within a gapped region as calculated based on the insert size. In comparison, BAUM selects reads over iterations by imposing more and more stringent uniqueness rules, thereby includes more reads for local assembly overall. Another example is IMAGE(Tsai et al. 2010). Unlike these two gap closure algorithms that use *k*-mers to extend contigs, we adopt the overlap-layout-consensus paradigm. The OLC methods played an important role in the DNA assembly using Sanger reads. However, the complexity of its pairwise alignment step is quadratic with respect to the number of reads. Besides, the short lengths of HTS reads make the significance of the overlap unreliable. Thus so far the OLC scheme has not been used in the assembly using HTS reads. By dividing the reads according to the neighboring unique genomic regions, the OLC is carried out locally around the end of each contig. Consequently, the computation complexity is not a problem at all and the overlapping significance is easier to be dealt with.

BAUM selectively merges adjacent contigs by statistically integrating their alignment result and the insert distance. Both alignment and insert distance have been used in the literature for gap closure. For example, IMAGE (Tsai et al. 2010) extends the two contigs for some iterations and merges them if the newly assembled contig aligns against both contigs and supposedly covers the gap. GapFiller (Boetzer and Pirovano 2012) does check the consistency between the alignment distance and insert distance by imposing a maximum on the difference. BAUM makes a further adjustment on the insert distances by fitting a linear regression on alignment distances. Since both distances could be mistaken, we take the highly robust approach of least trimmed squares. In the assembly shown in RESULTS, it turned out the slope estimate always centered around one while the intercept varied and could be as large as close to 100bp. In addition, we also provide a robust estimate of the standard error. In short, BAUM offers an empirical check of the consistency between the alignment distance and insert distance for contig merging, and it can be used in other DNA assemblers.

# METHODS

**UM reads and mapping criterion.** A key concept in the assembly method is the UM reads, which are defined by a mapping criterion and are obtained by a mapping algorithm. On the one hand, UM information is needed in three different working steps of the assembler: (i) generate layouts of UM reads; (ii) construct scaffolds by UM mate-pairs; (iii) extend contigs by reads whose mates are UM. On the other hand, the reference for mapping is modified along the assembly process. Thus the mapping criterion needs to be selected adaptively. Besides, the mapping speed is another important factor, since mapping is needed repeatedly. In the assembly shown in RESULTS, we adopted SEME to implement the mapping computation. Other than its high efficiency, the sensitivity of SEME can be evaluated approximately by a probabilistic model. The two main parameters in SEME are the lower bound of perfect match seed length $k$ and the upper bound of mismatches $m$. The probability that a read is mapped to a place on the reference genome by SEME can be calculated based on the following proposition (see Supplemental Methods S1 for proof).

**Proposition 1**: Suppose $r$ is an $l$-bp read and $s$ is an $l$-bp subsequence on the reference genome. Assume the events $\{r[i] = s[i]\}$, $1 \leq i \leq l$, are independent, and all base pairs have a common mismatch match rate $p$; here $r[i]$ and $s[i]$ represent the $i$-th base on $r$ and $s$ respectively. The probability that $r$ is able to be mapped to $s$ using SEME is:

$$\tau(k, m; l, p) = \sum_{n=0}^{m} \left[ \sum_{t=1}^{Q(l,n)} (-1)^{t-1} \binom{n+1}{t} \binom{l-kt}{n} \right] p^n (1-p)^{l-n},$$

where $Q(l, n) = \max\{t; l - kt \geq n\} \wedge (n + 1)$.

**Figure 3** shows the mapping rate versus mismatch rate under different mapping criteria calculated by the above formula. Under a loose criterion (small $k$ and large $m$ in SEME), the mapping rate is high but the read is easier to be mapped to a region with a relatively larger divergence, leading to multiple hits. While under a stringent criterion (large $k$ and small $m$), the mapping rate decreases but a successfully mapped read is more likely to be uniquely mapped.

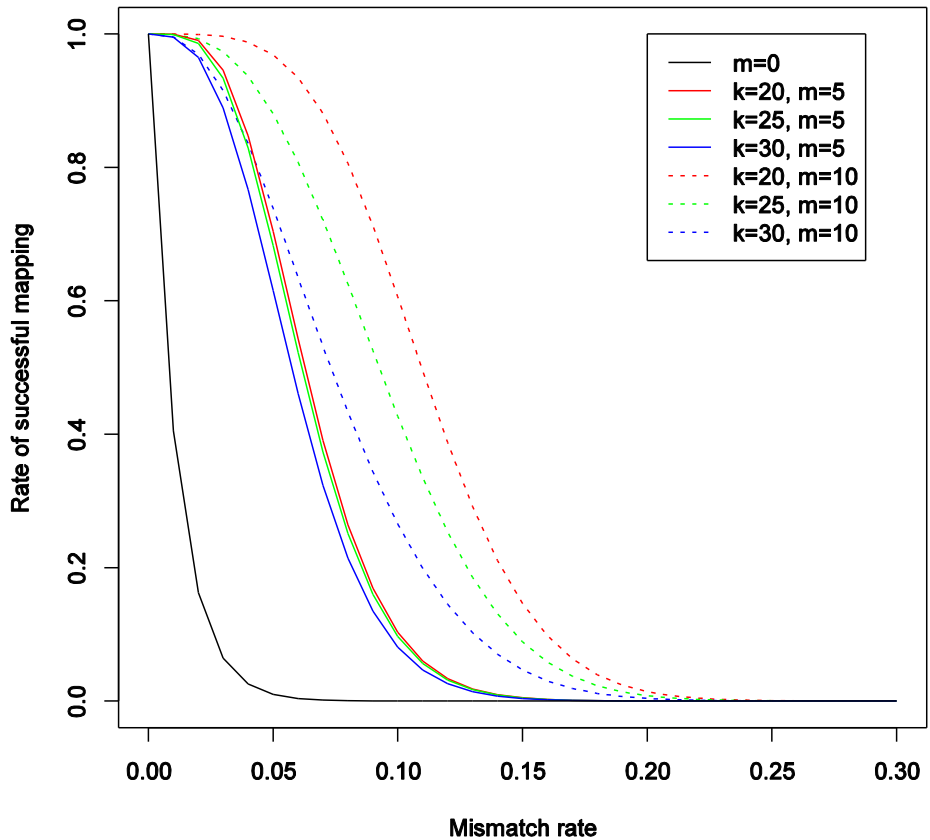

**Figure 3: The mapping sensitivity versus mismatch rate under different mapping criteria.**

A sequencing reads could be mapped to a homologous region other than its origin. If their nucleotide differences are approximately represented by a simple Bernoulli model, the mismatch rate is then a measure of divergence. Each curve depicts the mapping rate versus the mismatch rate under one criterion. Under a loose criterion (small $k$ and large $m$ in SEME), the mapping rate is high but the read is easier to be mapped to a region with a relatively larger divergence, leading to multiple hits. While under a stringent criterion (large $k$ and small $m$), the mapping rate decreases but a successfully mapped read is more likely to be uniquely mapped. These results obtained by Proposition 1 guide the selection of mapping criteria at different steps of the assembler BAUM.

At different working steps, the selection of criterion has different considerations. For the construction of layouts, if the inconsistency between the two genomes is relatively large, particularly in the initial mapping, we can set a loose criterion (small $k$ and large $m$) to optimize the mapping rate and the UM rate. As the reference gets closer and closer to the target, we tighten the criterion accordingly. To construct scaffolds and to extend contigs, we do impose a relatively stringent criterion to minimize false positives.

**Filtration.** In order to further reduce the uncertainty caused by repetitive elements, we apply filtration on the layout obtained above. In **Figure 4** we illustrate the ideas by three prototypical cases of homologous segments X, Y and Z located respectively on the reference and target genome. In case (a), under a certain mapping criterion, a read from Z can be mapped to both X and Y and is thus not a UM read. In case (b), a read from Z is possibly uniquely mapped to X but not to Y, so the UM criterion fails to filter it out. However, under a similar criterion, subsequences from Y of the same length as reads can be self-mapped to X. Thereby we propose a "self-mapping" strategy to filter out the layouts falling into this kind of non-unique regions. The "self-mapping" strategy consists of three steps: 1. generate all $l$-bp pseudo-reads from the reference genome; 2. map all pseudo-reads back to the reference genome under a criterion similar to the one used in initial mapping; 3. detect the regions covered only by UM pseudo-reads, namely, the unique regions on the reference genome. Then our proposed filtration removes all UM reads not contained in such regions from the layout. Finally, in case (c) shown in **Figure 4**, suppose that only the reads generated from X (or Y) can be mapped to Z, the depth in Z approximately follows a Poisson distribution with the intensity $l\lambda\tau(k,m;l,p)$, where $\lambda$ is the intensity of the Poisson distribution which is used to model the number of reads generated from each position on the target genome, and $\tau(k,m;l,p)$ is given by Proposition 1. If reads from both X and Y can be mapped to Z, then the depth will inflate. Consequently, filtration of positions whose depths exceed a proper threshold set by the depth distribution will eliminate this kind of uncertainty.

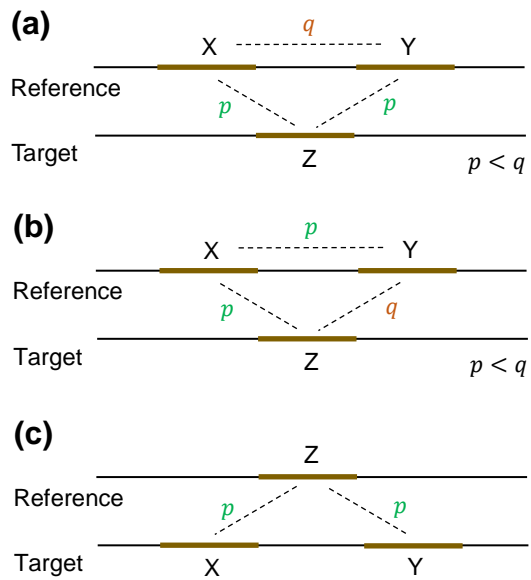

**Figure 4: Three prototypical cases of three homologous segments X, Y and Z locatedrespectively on the reference and target genome.** The mismatch rate between two copies is represented by p or q. (a) The divergence in terms of mismatch rate between X and Y is larger than those between the other two pairs, which are roughly the same; (b) The divergence between Z and X is similar to that between X and Y, and is smaller than that between Z and Y; (c) X and Y on the target have only one homologous region Z on the reference. Under a certain mapping criterion, a read from Z can be mapped to both X and Y in case (a) and is thus not a UM read. In case (b), under a similar criterion, a read from Y can be self-mapped to X along with Y, and eventually X and Y will be marked as non-unique regions. In case (c), reads generated from both X and Y can be uniquely mapped to the same region Z. Then Z will be detected by a change in depth, and thereby be filtered out.

**Layout split.** A collection of layouts are generated after mapping and filtration. We notice that some layouts need to be split at the sites of structural variations (SV) between the two genomes. Shown in **Figure 5(a)** is an indel example in which the layouts corresponding to two separated regions on the target genome are unexpectedly connected after mapping. In order to avoid mis-assembly at the structural level, it is necessary to detect all the breakpoints of SV within each layout. It is noted that after unique mapping under a selected criterion and filtration, the relevance of layout split is primarily on the unique regions.

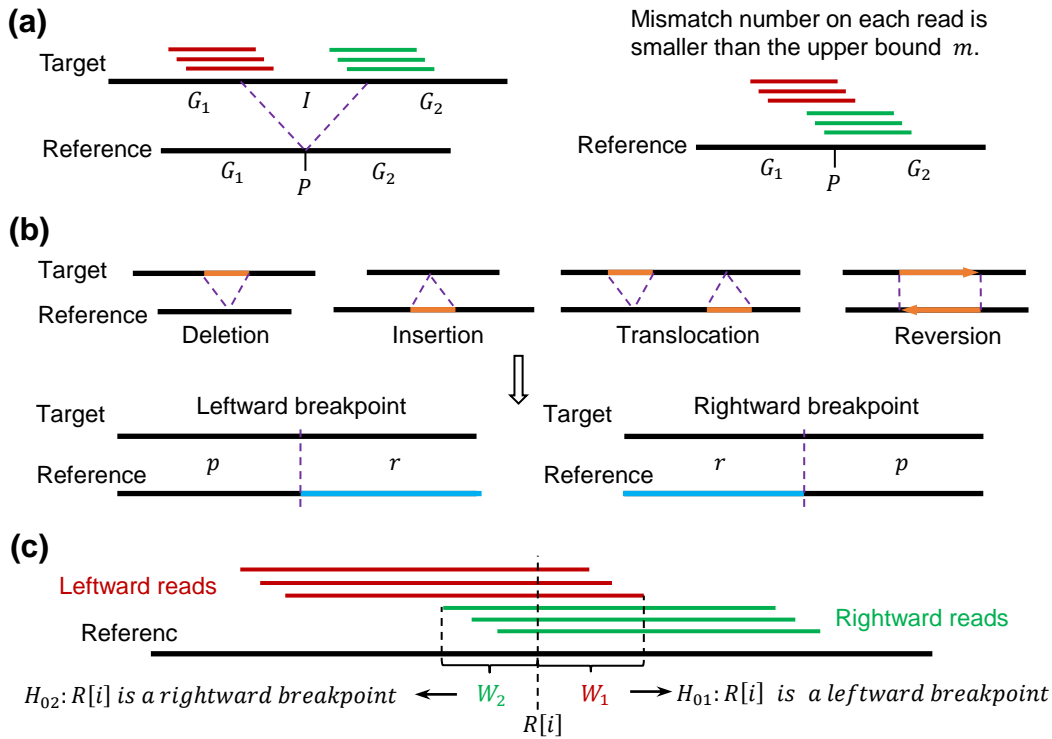

**Figure 5: Illustration of layout split.** (a) Region $I$ between $G_1$ and $G_2$ on the target genome is deleted on the reference genome, causing the reads generated from two separated regions merge into one layout after mapping. (b) The occasions of SV breakpoints can be decomposed

as two simplified cases: leftward breakpoint and rightward breakpoint. For the leftward (rightward) breakpoint, the mismatch rate of the nucleotides is $p$ on the left (right) and $r$ on the right (left). $p$ is the mismatch rate of common region; $r$ is the mismatch rate under a random occasion, and $p < r$. (c) The base $R[i]$ on the reference genome divides the reads covering it into two classes: leftward reads (the left parts are longer than the right parts) and rightward reads (the right parts are longer than the left parts). At each position, $W_1$ and $W_2$ are the maximum tail length of the leftward reads and rightward reads. The null hypothesis regarding the leftward breakpoint and rightward breakpoint are tested by the statistics $W_1$ and $W_2$ respectively. The layout is split at $R[i]$ if one of the tests fails to be rejected.

In the context of layout split, we simplify various kinds of SVs into two cases (**Figure 5(b)**): leftward breakpoint and rightward breakpoint. Our statistical model for the leftward (rightward) breakpoint is as follows: the mismatch rate of the nucleotides is $p$ on the left (right) and $r$ on the right (left), where $p$ is the mismatch rate of common region, $r$ is the mismatch rate under a random setting, and $p < r$. Then the detection of breakpoints can be formulated by two hypothesis tests at each nucleotide base $R[i]$ on the reference genome (**Figure 5(c)**). That is, the null hypotheses are $H_{01}$: $R[i]\ is\ a\ leftward\ breakpoint$, and
$H_{02}$: $R[i]\ is\ a\ rightward\ breakpoint$. Here $R[i]$ represents the $i$-th base on the reference genome. The layout is split at $R[i]$ if one of the tests fails to be rejected. We need the following definition for the test statistics.

**Definition**: Consider a read covering a base on the reference genome. If its left part is longer than the right part, it is called a leftward read; otherwise, it is called a rightward read. The shorter part of a read is called its tail.

Heuristically, we use the statistics $W_1$ and $W_2$, which are defined as the maximum tail length of the leftward reads and rightward reads at $R[i]$, for the two tests respectively (**Figure 5(c)**). For the sake of simplicity, we assume that reads are generated from the genome according to a standard Poisson process with intensity $\lambda$. The distribution of the test statistics can be evaluated by the following.

**Proposition 2**: Under $H_{01}$, $\forall t \in \mathbb{N}$, $1 \leq t \leq \lfloor l/2 \rfloor$, $Pr(T_1 \geq t) \leq 1 - \left[ \exp\left(-\lambda \sum_{n=t+1}^{\lfloor l/2 \rfloor} f(m|n,r)\right) + \exp\left(-\lambda \sum_{n=\lceil l/2 \rceil}^{l-1} f(m|n,r)\right) - 1 \right]$,

where $f(\cdot | n, p_r)$ is the cumulative distribution function of the binomial distribution $B(n,r)$. Similar result can be obtained for $W_2$ under $H_{02}$. See Supplemental Methods S1 for proof.

Given a threshold $\alpha$ on type I errors, we can choose the rejection region according to the above probability. We note that the primary control is the rate of false positives in which necessary splits are missed. Although the over-protection may lead to more false splits, they can be saved in the scaffolding and extension steps.

**Scaffolding and contig extension.** After the step of layout split, we obtain the initial contigs and move on to the loop on the right in **Figure 1**. At this point, we should have the mapping information of all the reads with respect to the initial contigs, particularly, the UM reads. We can use the mapping information of UM reads to carry out two tasks. On the one hand, scaffolds are built based on the mapping coordinates of the mate-pairs and their library DNA sizes. On the other hand, the reads whose mates are uniquely mapped to the end of each contig are collected for contig extension. Although these collected reads may overlap with repetitive elements, the uncertainty has already been resolved by localization.

**Local overlap-layout-consensus assembly and computation complexity.** The overlap-layout-consensus scheme is a time-proven assembly method for Sanger sequencing reads. A typical OLC example is PHRAP, whose complexity is quadratic with respect to the number of reads. As the scale of reads goes up dramatically in the HTS technology, OLC is no longer capable of handling sequencing reads altogether. In comparison, BAUM applies OLC to only a relatively small subset of reads in each contig extension. As these reads' mates are uniquely mapped to the end of the contig, each subset is practically from a local region of the genome, and we term it as the local OLC. Since the extensions of all contigs are independent of each other, we can implement local OLC through parallel computation. The time complexity is given as follows.

**Proposition 3**: Denote the number of contigs by $N_C$ and the sequencing coverage by $D$. Then the average time complexity of local OLC for contig extension is $O(D^2 l^2 N_C)$.

Namely, the time complexity is linear with respect to the number of contigs, and is quadratic only with respect to the read depth.

**Gap distance between adjacent contigs and merging of adjacent extended contigs.** The scaffolding procedure takes the information of mate pair and library DNA sizes as input. Its output includes estimates of the adjacency information of all contigs as well as the gap distances between adjacent contigs. Although the estimates are not accurate, they serve as hypotheses in the later contig merging.

The first step of contig merging is the application of local alignments to adjacent extended contigs. The existence of high scoring segments in an alignment suggests a possible overlap of the two adjacent contigs. However, the decision of contig overlap and merging is not simple but subtle. One challenge is the complicated patterns observed in the local alignment. Three typical patterns are illustrated in Fig 2(a), 2(b) and 2(c). In the first pattern, the right end of one extended contig is perfectly aligned to the left end of the other contig, and we refer to it as "no hanging end". In the second pattern, one end of an extended contig is well aligned to a segment inside the other contig, and we refer to it as "one hanging end". In the third pattern, the highest scoring segment corresponds to portions inside both contigs, and we refer to it as "two hanging ends". Even though the no-hanging-end pattern seems to provide stronger evidence of contig overlap, we could not simply reject the other two patterns because of possible heterozygous DNA content including insertions/deletions in a diploid genome.

Here we report one statistical approach for contig merging. For the sake of clarity, we introduce some notations. Let $a, b$ be two adjacent contigs determined by the scaffolding procedure. Their extended contigs are denoted by $\tilde{a}, \tilde{b}$, and the two extended lengths are denoted respectively by $ext(a)\ and\ ext(b)$. We first consider the case of "no hanging end". Suppose that the two adjacent extended contigs $\tilde{a}, \tilde{b}$ significantly overlap by $o(\tilde{a}, \tilde{b})$ base pairs from the result in local alignment. Then we can estimate the alignment distance between $a, b$ by $r(a,b) = ext(a) + ext(b) - o(\tilde{a}, \tilde{b})$. On the other hand, the scaffolding procedure also estimates the same gap distance using the insert sizes of mate pairs that are uniquely mapped to the neighboring contigs $a, b$, and we denote this insert distance by $s(a,b)$. The alignment and insert distance $r(a,b)$ and $s(a,b)$ are obtained from independent sources of information. If the difference $|r(a,b) - s(a,b)|$ is small, then we merge the two extended contigs $\tilde{a}, \tilde{b}$. Statistically, we pool $\{r(a,b),\ s(a,b)\}$ for all the cases of "no hanging end", and fit a simple linear regression line to the bivariate data. However, a fraction of $\{r(a,b),\ s(a,b)\}$ could deviate from the line because either $r(a,b)$ or $s(a,b)$ could be mistaken. Thus we need a robust regression method that would not break down in the presence of a possibly large fraction of outliers. In this study, we take the least trimmed squares approach. The robust estimate of the slope should be close to 1; otherwise it indicates some systematic bias occurs in the gap distance estimates. Once the robust regression line is fitted, the insert distances are adjusted by the "good" alignment results corresponding to the LTS subset, and the residuals can be calculated for the consistency check.

**Iterative assembly** The current assembly result can be taken as the reference in the next iteration of mapping-scaffolding/extension-merging. As contigs are extended, some may gradually overlap with the repetitive elements. After some iterations we cannot obtain any more UM reads if we do not change the mapping criterion. Nevertheless, if we apply a more stringent mapping criterion at this point, more UM can be obtained and consequently further extension can be gained.

**Software availability.** The source code is available at https://github.com/Zhanyu-Wang-AMSS/BAUM.

# DATA ACCESS

The raw reads of sequencing data, including seven Illumina HiSeq paired-end/mate-pair libraries (insert size 300, 400, 1K, 2K, 5K, 10K and 20K) and 454 long reads, are available in SRA database of NCBI under the accession numbers SRX1156187, SRX1156186 and SRX1156057 (Zhang et al. 2015).

# ACKNOWLEDGEMENTS

We thank Wen Wang at Kunming Institute of Zoology, Chinese Academy of Sciences (CAS) for providing us with the Illumina HiSeq and Roche 454 sequencing data of *Oryza longistaminata*. This work was supported by the Strategic Priority Research Program of the Chinese Academy of Sciences (Grant No. XDB13040600), the National Natural Science Foundation of China (Grant No. 91530105，91130008), the National Center for Mathematics and Interdisciplinary Sciences of the CAS, and the Key Laboratory of Systems and Control of the CAS.# DISCLOSURE DECLARATION

None declared.